# Evolución y uso del vocabulario de la ciencia de datos. ¿Cuánto hemos cambiado en 13 años?

Igor Barahona.

Laboratorio de Aplicaciones de las Matemáticas. Unidad Cuernavaca del Instituto de Matemáticas. UNAM

igor@im.unam.mx

Palabras clave: Ciencia de datos, minería de textos, estadística multivariante, análisis de contenido

## Resumen

Aquí presento una investigación sobre la evolución y uso del vocabulario en la ciencia de datos en los últimos 13 años. Partiendo de un análisis estadístico riguroso, se analiza una base de datos con 12,787 documentos que contienen las palabras “data science” en el título, resumen o palabras-clave. Se propone clasificar la evolución de esta disciplina en tres periodos: surgimiento, crecimiento y auge. Palabras características y documentos pioneros son identificados para cada periodo. Al proponer el vocabulario distintivo y temas relevantes de la ciencia de datos y clasificados en periodos de tiempo, estos resultados agregan valor a la comunidad científica de esta disciplina.

## Introducción

El número de personas, empresas y organizaciones que están “siempre conectadas” al Internet creció exponencialmente durante la última década. Este fenómeno ha generado enormes cantidades de datos, y también dio lugar al surgimiento del término "big-data". De acuerdo con SAS (2022), el término "big-data" es un anglicismo utilizado para describir un gran volumen de datos, tanto estructurados como no estructurados. Las sociedades contemporáneas se caracterizan por ser potentes generadoras de "big-data". La ley de Moore, que establece que el número de transistores incluidos en un circuito integrado se duplica cada dos años (Moore, 1965), describe adecuadamente el crecimiento en el número de dispositivos conectados a Internet durante la última década. De acuerdo con Van der Aalst (2016) otros fenómenos como la capacidad de almacenamiento digital y el número de pixeles por unidad de área también han tenido un crecimiento exponencial. Es posible describir el comportamiento de estos fenómenos en términos de la Ley de Moore. Lo anterior genera un panorama abundante en datos el cual exige metodologías novedosas, con la finalidad de transformar tales datos en información útil para los procesos de toma de decisiones. El IDC (2021) menciona que el universo digital se está expandiendo de forma similar al universo físico, duplicando su tamaño cada dos años. Se estima que el universo digital superó los 50 zetabites ($50x10^{12}$ de GB) al final del año 2021.

En medio de esta transición de sociedades analógicas a otras digitales, surge una disciplina denominada "Ciencia de Datos". Las sociedades digitales contemporáneas enfrentan nuevos desafíos, los cuales no pueden ser resueltos a través de los métodos científicos tradicionales. Por ejemplo, no es posible extraer información útil de grandes cúmulos de datos tomando como punto de partida los métodos matemáticos tradicionales bien conocidos. Para hacer investigación pertinente a partir de grandes volúmenes de datos, se requiere la combinación de las ciencias matemáticas con las computacionales en torno a un dominio específico de aplicación. De esta forma, la ciencia de datos surge como una fusión entre las ciencias computacionales, las matemáticas y la estadística en torno un área específica de interés. Skiena (2017) afirma que la ciencia de datos se encuentra en la intersección de las ciencias computacionales, la estadística y las matemáticas, las cuales establecen sinergias para investigar un fenómeno en particular.

En este contexto, investigar el uso y evolución del vocabulario en el área de la ciencia de datos se establece como el objetivo general para este trabajo. Cinco partes componen este documento. En la siguiente sección se presenta una revisión de literatura con los aspectos más remarcables de esta disciplina y algunas de sus definiciones formales. Las características de la base de datos analizada, así como la metodología utilizada, aparecen en la tercera sección. Los resultados obtenidos se dejan en cuarto lugar. Las conclusiones y las futuras líneas de investigación aparecen en la última sección.

## Revisión de literatura

Autores como Donoho, (2017) y Cao, (2017) coinciden en que Tukey (1962), en su distinguido artículo titulado "El futuro del análisis de datos", introduce una de las primeras definiciones formales del concepto "ciencia de datos" de la siguiente manera: *la estadística debe complementarse, entre otras cosas, con procedimientos para analizar datos, técnicas para interpretar los resultados de tales procedimientos, formas de planificar la recolección de los datos con la finalidad de hacer su análisis más fácil, más preciso y más exacto, además de aplicar toda la maquinaria estadística (matemática) al análisis de datos*.

Peter Naur, en su libro titulado "Concise Survey of Computer Methods", utiliza el término ciencia datos como equivalente en significado a las ciencias computacionales, con el propósito de ilustrar la relación entre las ciencias naturales y el uso de los datos (Belzer, 1976). En el año 1998, durante la reunión anual de la Asociación Internacional de Sociedades de Clasificación, Hayashi (1998) introduce el concepto de "Dēta no bunseki-teki chōsa", que se puede traducir del japonés al español como la tarea de realizar una exploración analítica de los datos. Hayashi define a la ciencia de datos como una disciplina que permite clarificar o entender un fenómeno mediante el análisis de datos y de experimentos cuidadosamente diseñados. Jeff Wu, profesor de la Escuela de Ingeniería Industrial y Sistemas de Ingeniería del Tecnológico de Georgia, sugirió que la estadística podría ser renombrada como ciencia de datos, y que esta última debería hacer énfasis en la recopilación de datos, el modelado matemático y los sistemas informáticos (Wu, 1997). Por otra parte, el artículo titulado "Ciencia de datos: un plan de acción para expandir las áreas técnicas del campo de la estadística" y escrito por Cleveland (2007), propone que la ciencia de datos es la nueva cara de la estadística moderna, al incorporar la informática, y por lo tanto, mejorar la precisión en el análisis de datos. Adicionalmente, el autor propone un plan curricular para el desarrollo de la ciencia de datos, el cual es susceptible de ser aplicado en gobiernos, laboratorios, organizaciones y universidades. En 2002 aparece el "Journal of Data Science", el cual fue una de las primeras revistas científicas que aparecieron y especializadas en esta área. Le siguieron revistas como "International Journal of Data Science and Analytics", "Data Science and Management" y "Data Science and Engineering", entre otras.

En el año 2012, la ciencia de datos fue reconocida como una de las áreas laborales con mayor crecimiento y expansión. De igual forma, el trabajo de "científico de datos" fue resaltado como uno de los mejores remunerados en Estados Unidos (Davenport & Patil, 2012). Estos autores destacan las cualidades y responsabilidades que deben poseer los profesionales dedicados a la ciencia de datos, así como su importancia en las organizaciones y entidades de gobierno. En una encuesta distribuida entre 11,514 miembros de la Sociedad Americana de Estadística en el año 2020, un 82% de los encuestados manifestaron estar laborando en las áreas como la estadística aplicada y la ciencia de datos. Un 92% afirmó que, si un alumno o joven les pidiera un consejo sobre como seleccionar una carrera universitaria, le animarían a estudiar estadística o ciencia de datos. Por el contrario, solo un 7% mencionó que desalentarían a los alumnos o jóvenes estudiar una carrera relacionada con estas áreas (ASA, 2020).

Derivado de la pandemia ocasionada por la propagación del COVID-19 a nivel mundial, la interacción a través de medios digitales, así como el número de dispositivos "siempre contactados" a internet, crecieron de forma importante. Lo anterior acentuó aún más las tendencias antes descritas, relativas a la generación, recolección, almacenamiento y análisis de datos. Tomando en cuenta este escenario creciente en datos, así como un auge en las sociedades digitales, la ciencia de datos tiene una perspectiva de crecimiento para los próximos años pospandemia.

## Metodología

Con la finalidad de lograr el objetivo de esta investigación, se propone una metodología de cinco pasos. Iniciando con una explicación de los procedimientos relativos a la recolección de los datos, pasando por la presentación de las herramientas computacionales y cerrando con los métodos estadísticos aplicados en el análisis.

**I.** *La base de datos*. Se realizó una búsqueda en la librería digital Scopus (2022) con los siguientes criterios de búsqueda: todos los documentos deben incluir la frase "data science", ya sea en el título, resumen o palabras claves. Este criterio de búsqueda arroja un total de 14,162 documentos. Del conjunto anterior, un total de 945 documentos fueron descartados porque no cubrían las características para ser considerados documentos de investigación, como por ejemplo cartas al editor o notas aclaratorias. Adicionalmente, 430 documentos carentes de resumen también fueron removidos. La base de datos resultante y utilizada en el análisis contiene 12,787 documentos que comprenden el periodo entre los años 2009 y 2022. En promedio cada documento (abstract) tiene 168 palabras, de las cuales 105 son palabras únicas. De esta forma, el porcentaje de términos únicos es igual a 65% para cada documento en promedio.

**II.** *Herramientas computacionales*. Los análisis se realizaron bajo el ambiente de programación R, versión 4.1.2 (Bird-Hippie) liberada para su descarga en noviembre de 2021. Se utilizaron las siguientes librerías especializadas en minería de textos “tidytext”, “factoextra”, “topicmodels”, “FactoMiner” y “wordCloud”, entre otras (CRAN, 2022).

**III.** *Análisis estadístico descriptivo básico*. Para la primera parte del análisis se presentan gráficos de barras con las palabras más frecuentes, gráfico de barras para el número de publicaciones por año y gráfico de barras con los tipos de publicaciones. Así también en esta etapa se removieron los “stopwords”, las cuales consisten principalmente en artículos, pronombres y preposiciones, entre otras. El algoritmo propuesto por Schofield, Magnusson & Mimno (2017) fue adaptado para este propósito.

**IV**. *Análisis estadístico multivariante*. El Análisis de Correspondencias (AC) es un método estadístico de reducción de dimensionalidad propuesto por Jean-Paul Benzécri en la década de los 60s, el cual ha sido ampliamente utilizado para investigar conjuntos de datos de gran complejidad o tamaño. El punto de partida para realizar un AC es una matriz documentos palabras, (denotada como **X**), con dimensiones $n$ x $p$, donde $n$ representa el número de observaciones (renglones) y $p$ representa las variables (columnas). En la Figura 1, una representación de la matriz **X**.

$$
\begin{array}{c|cccc|c}
 & \mathrm{B}_1 & \mathrm{B}_2 & \cdots & \mathrm{B}_J & \\
\hline
\mathrm{A}_1 & f_{11} & f_{12} & \cdots & f_{1J} & f_{1\cdot} \\
\mathrm{A}_2 & f_{21} & f_{22} & \cdots & f_{2J} & f_{2\cdot} \\
\vdots & \vdots & \vdots & \ddots & \vdots & \vdots \\
\mathrm{A}_I & f_{I1} & f_{I2} & \cdots & f_{IJ} & f_{I\cdot} \\
\hline
 & f_{\cdot 1} & f_{\cdot 2} & \cdots & f_{\cdot J} & n
\end{array}
$$

**Figura 1**. Estructura de una matriz documentos-palabras (MDP)

Para la Matriz Documentos Palabras (MDP), $f_i = \sum_j f_{ij}$ representa la frecuencia marginal de $\mathrm{A}_I$. Por otra parte, $f_j = \sum_i f_{ij}$ representa la frecuencia marginal para $\mathrm{B}_J$. Tomar en cuenta que para cada categoría $\mathrm{A}_1, \dots, \mathrm{A}_I$, le corresponden los valores contenidos en el vector $\boldsymbol{a} = (a_1, \dots, a_I)$. A su vez, cada elemento de $\boldsymbol{a}$ está dado por la frecuencia absoluta en la respectiva posición de la MDP. De forma similar para las columnas $\mathrm{B}_1, \dots, \mathrm{B}_J$, le corresponden los valores del vector representado por $\boldsymbol{b} = (b_1, \dots, b_J)$.

Sean $U$ y $V$ variables compuestas, las cuales se definen como $U = \boldsymbol{X}a$ y $V = \boldsymbol{Y}b$, respectivamente. Entonces el análisis de correspondencias consiste en encontrar la combinación lineal para los vectores $\boldsymbol{a}$ y $\boldsymbol{b}$, los cuales maximizan la correlación entre las variables compuestas $U$ y $V$. La Descomposición en Valores Singulares (DVS) para la matriz **X** se puede expresar de la siguiente forma.

$$\mathbf{D}_a^{-\frac{1}{2}}(\mathbf{X} - ab')\mathbf{D}_b^{-\frac{1}{2}} = U\mathbf{D}_\lambda V' \tag{1}$$

En el contexto de la minería de textos y el análisis de contenido, la metodología previamente descrita hace posible cuantificar relaciones entre palabras y documentos, en función de su connotación semántica. Lo anterior a través de la comparación de las *categorías-renglón* por una parte, y las *categorías-columna* por la otra, tomando como referencia los valores singulares obtenidos con la formulación (1). Dos o más palabras con valores singulares cercanos tendrán un significado semántico semejante. Lo mismo aplica para los documentos con valores singulares cercanos, de los cuales se infiere que están utilizando vocabulario semejante, o bien tocando temas similares. Al introducir la variable año de publicación como suplementaria al análisis, es posible identificar la evolución del vocabulario en el tiempo. El año de publicación estará caracterizado por las palabras o documentos, que tengan valores singulares cercanos. Posteriormente, a través de las matrices obtenidas con la DVS, es posible generar visualizaciones denominadas coloquialmente como “nubes de palabras” o “nubes de documentos”, las cuales nos permiten identificar la evolución del vocabulario en el tiempo, de una forma sencilla e intuitiva.

Considerando que esta metodología está basada principalmente en principios matemáticos, es posible replicarla para cualquier tema, idioma o contexto. En la siguiente sección, los resultados obtenidos a través de la aplicación de esta metodología se presentan en un caso de estudio, el cual ilustra la evolución del vocabulario en el tema de la ciencia de datos.

## Resultados

**I.** *Análisis estadístico descriptivo básico*. El número de publicaciones por año para el periodo del año 2009 al 2022 se presenta en la Figura 2. Para estimar el número de publicaciones con las cuales cerrará el año 2022, así como las esperadas para el 2023, se ajustó el siguiente modelo de segundo orden: $y(x) = 46.9x^2 - 370.1x + 579$, a través del método de los mínimos cuadrados. Con un valor para el coeficiente de ajuste $R^2 = 0.98$, se estima que el número de publicaciones será de 4,590 y 5,580 para los años 2022 y 2023 respectivamente.

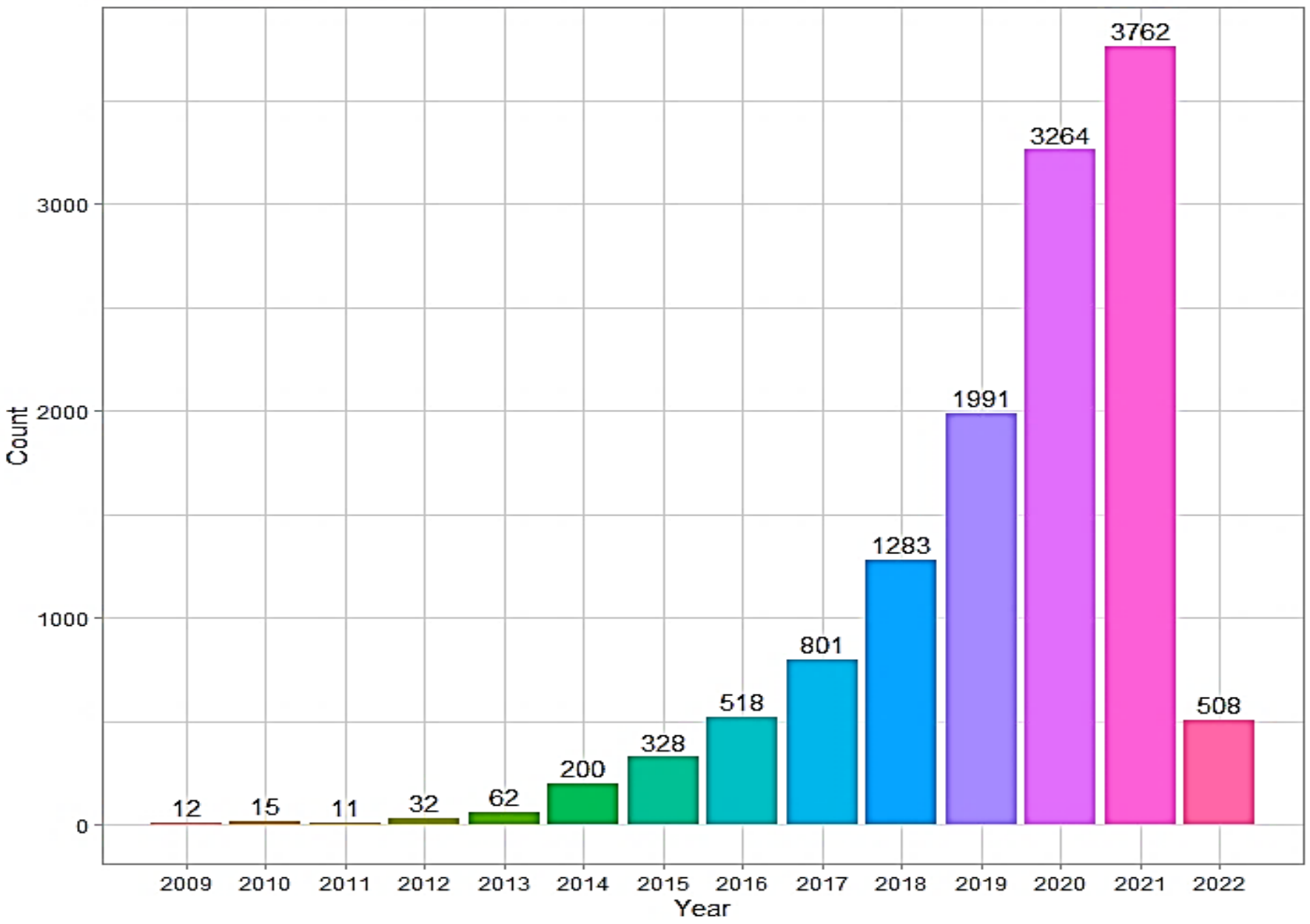

**Figura 2**. Publicaciones por año en el tema de Ciencia de Datos

Con la finalidad de hacer una primera aproximación al uso del vocabulario en el tema de ciencia de datos, se realizó un gráfico de barras para representar las 30 palabras con mayor frecuencia.

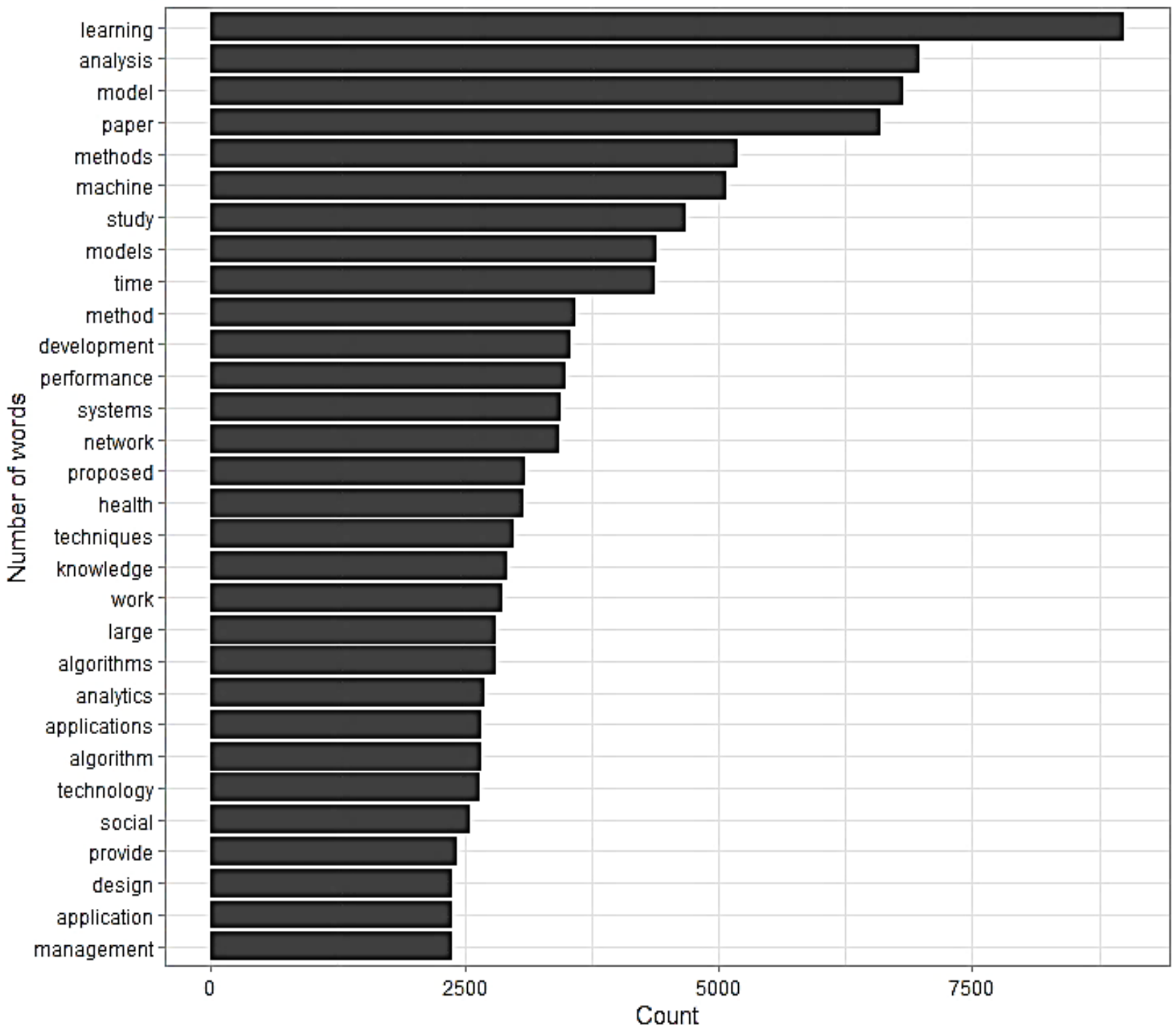

**Figura 3**. Palabras más frecuentes en el tema de ciencia de datos. Periodo del año 2009 al 2022

Palabras como *learning* (8,996), *analysis* (6,980), *model* (6,818), *paper* (6,591), *methods* (5,186) y *machine* (5,087) aparecen como las más frecuentes en la base de datos objeto de estudio. A través de esta primera aproximación, podemos tener una idea general de algunos de los temas predominantes para la ciencia de datos durante la última década, como por ejemplo el “aprendizaje de máquina” o “métodos de aprendizaje de máquina”. Las treinta palabras que aparecen en la Figura 1 y sus respectivas frecuencias representan el 10.1% del volumen total del vocabulario.

En cuanto al tipo de publicación, el 50% del total de las publicaciones en este tema corresponden a memorias de congreso. La otra mitad se divide entre artículos (37%), revisiones de artículo (7%), capítulos de libro (3%), revisiones de memorias de congreso (2%) y libros (1%). De esta forma, las memorias de congreso es el tipo de publicación predilecto para los académicos, científicos y profesionales que realizan investigación en temas relacionados con la ciencia de datos.

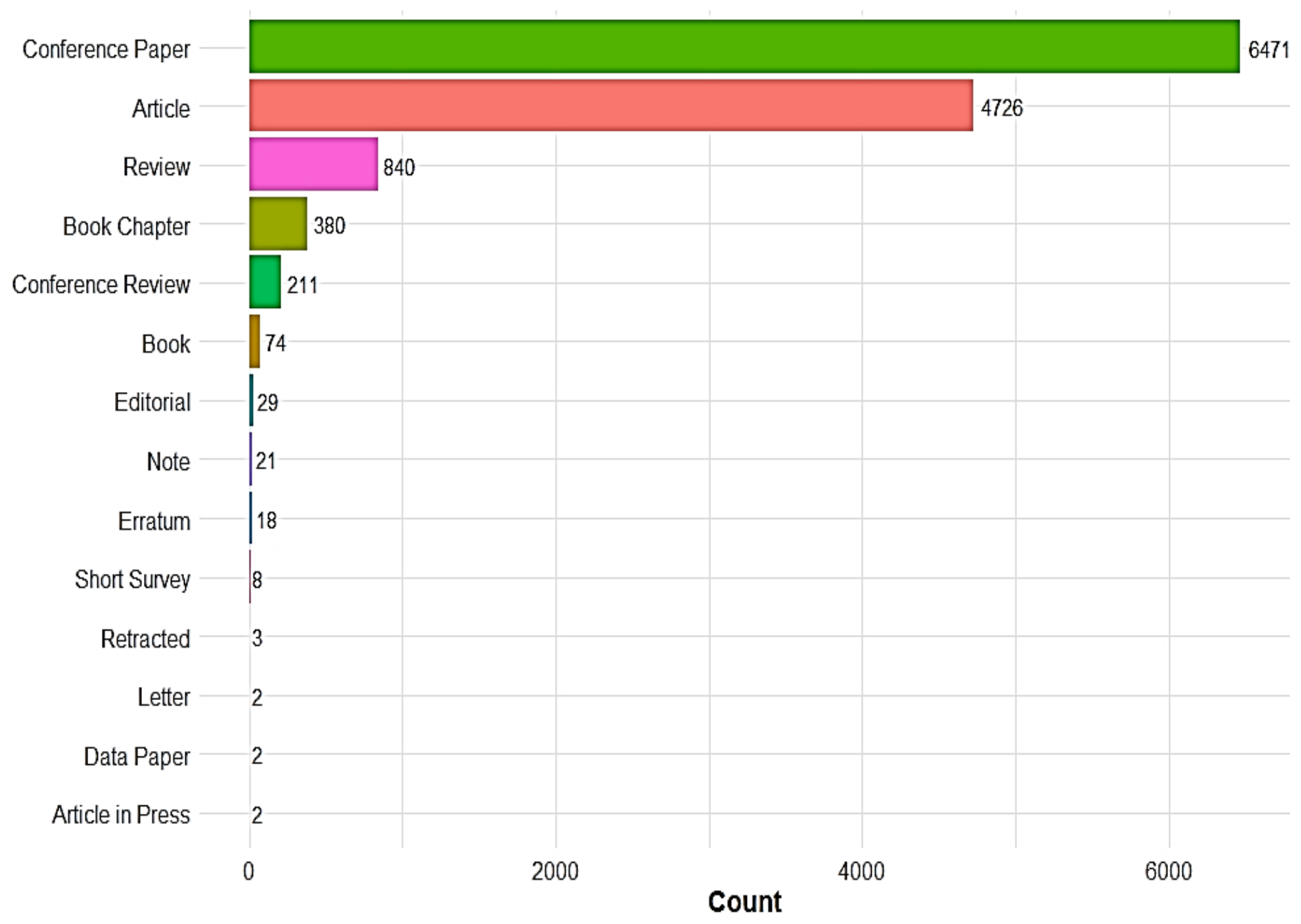

**Figura 4**. Tipo de publicación para el tema "ciencia de datos". Periodo del año 2009 al 2022

Los artículos publicados en memorias de congreso generalmente son el resultado de la participación en una reunión anual de un gremio o asociación. Usualmente, estos documentos son cortos, precisos y con un número reducido de páginas. Posteriormente, una selección de estos documentos es publicada, la cual es denominada memorias de congreso. Por otro lado, los artículos son publicados generalmente en revistas especializadas en un tema en particular. A diferencia de las memorias de congreso, los artículos de revista pueden ser publicados sin el requerimiento de hacer una presentación oral del trabajo. En general, los artículos publicados en revistas reconocidas pueden tener mayor impacto que las memorias de congreso. Sin embargo, lo anterior puede ser diferente en función de cada área de investigación en particular.

**II.** *Análisis estadístico multivariante (evolución y uso del vocabulario en el tiempo)*. En la Figura 5 aparece una representación de la evolución y uso del vocabulario en el tiempo para el área de ciencia de datos. Las posiciones de las palabras y los años están dadas por los valores singulares en el primer y segundo componente, los cuales son empleados como coordenadas cartesianas del tipo ($x$, $y$). Para esta disciplina se identifican tres periodos en función de los 12,787 documentos investigados. El primero denominado "*Surgimiento*" y que está comprendido del año 2009 al 2012. El segundo caracterizado por un notable

"*Crecimiento*", se ubica entre los años 2013 y 2018. Finalmente, el tercer periodo de "*Auge*" (Boom) se contextualiza entre el 2019 y 2022.

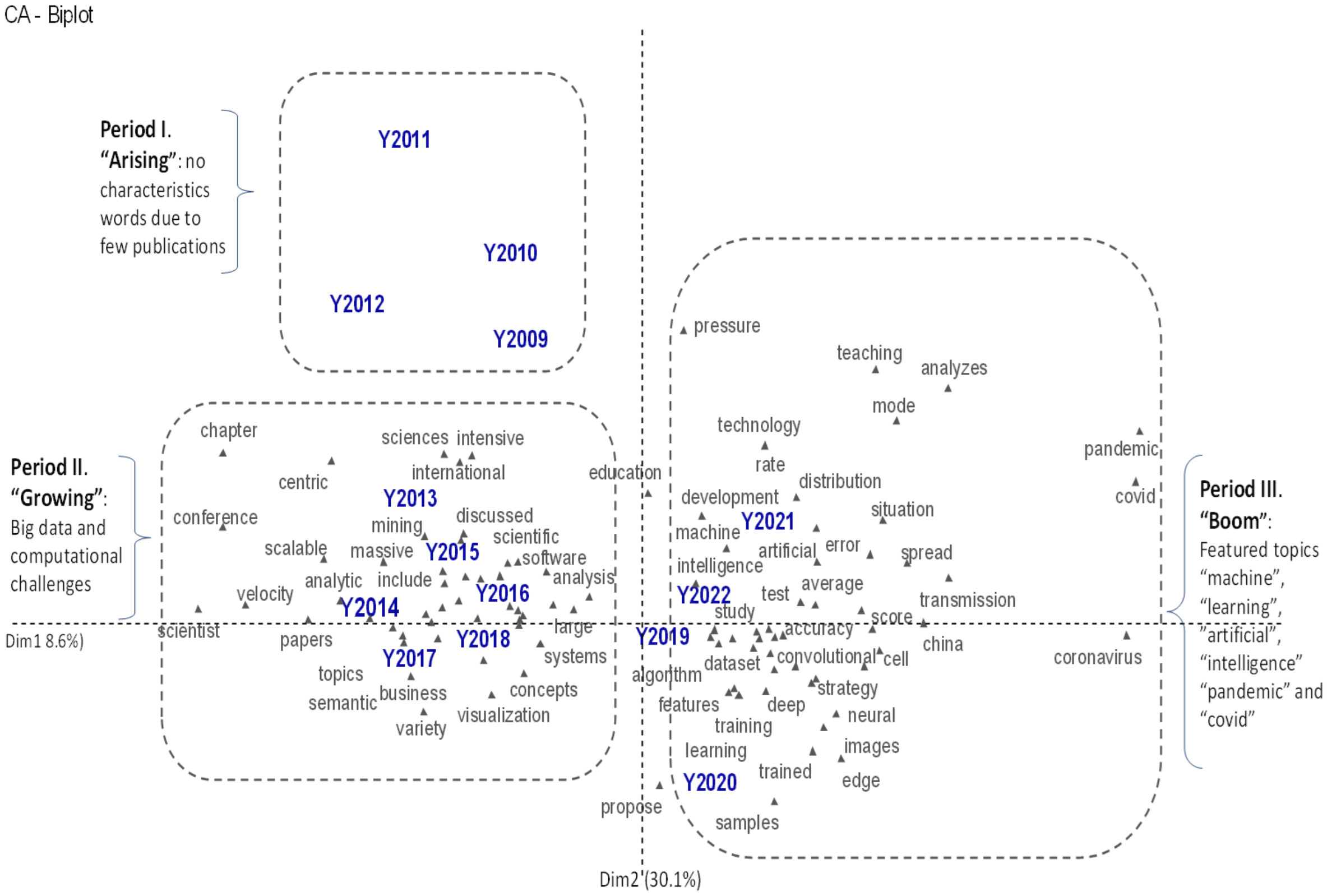

**Figura 5**. La evolución y uso del vocabulario en el área de la ciencia de datos

La ausencia de palabras características en el primer periodo ("Surgimiento") se debe principalmente al bajo número de documentos que corresponden a estos cuatro años. 70 artículos fueron publicados entre 2009 y 2012, lo que representa un 0.005% del total de los documentos investigados. Sobre estos resultados, se puede inferir que durante en este primer periodo se realizó investigación en el área, la cual fue publicada en periodos posteriores.

El segundo periodo de "Crecimiento", comprendido entre los años 2013 y 2019, se caracteriza por el uso de palabras como "minería", "masivo", "escalable", "velocidad", "software", "grande", "sistemas", "visualización" y "análisis", entre otras. De lo anterior se infiere que para este periodo los temas de investigación relevantes estaban relacionados con el análisis de grandes cúmulos de información (big-data), así como los retos computacionales que implica analizar satisfactoriamente estos datos masivos, como por ejemplo la escalabilidad y la velocidad. Algunos de los artículos característicos de este periodo son: *Process mining: Data science in action* del año 2016 y con 1,164 citas, *Data science, predictive analytics, and big data: A revolution that will transform supply chain design and management* del año 2013 y con 710 citas, y *Big data: Astronomical or genomical?* del año 2015 y con 668 citas. Debido limitaciones de espacio, se omite mencionar otros artículos característicos de este periodo.

El tercer periodo de "Auge" y comprendido entre los años 2019 y 2022 se caracteriza por términos como "máquina", "aprendizaje", "precisión", "entrenamiento", "convolucional", "estrategia", "artificial", "inteligencia", "Covid" y "pandemia". De esta forma, los temas de investigación distintivos de este periodo se relacionan con las aplicaciones del aprendizaje de máquina, la inteligencia artificial, las redes neuronales, la precisión de los algoritmos desarrollados durante el periodo, así como sus aplicaciones en torno a la pandemia y la propagación del COVID-19. Un total de 9,525 artículos fueron publicados entre los años 2019 y 2022, los cuales representan el 75% del total de la base de datos. Por lo tanto, en este periodo se introduce la mayor parte del vocabulario representativo de la ciencia de datos. Algunos de los artículos más representativos de este periodo son: *Explaining explanations: An overview of interpretability of machine learning*, con 444 citas y publicado en el año 2019, *The promise of artificial intelligence in chemical engineering: Is it here, finally?* con 207 citas y publicado en 2019, y *Artificial intelligence vs COVID-19: limitations, constraints and pitfalls* con 117 citas y publicado en 2020.

## Conclusiones

En este trabajo se investiga una base de datos con 12,787 documentos, con la finalidad de hacer conclusiones relevantes sobre la evolución y uso del vocabulario de la ciencia de datos durante los últimos 13 años. Los resultados identifican tres periodos. El primer periodo de "Surgimiento" (2009-2012), se caracteriza por la ausencia de palabras clave. Se infiere que durante tal periodo se realizaron todas las investigaciones que fueron publicadas en años posteriores. Entre los temas distintivos para el segundo periodo de "Crecimiento" (2013-2018) se encuentran el análisis de grandes cúmulos de información (big-data) y los retos computacionales que implica analizar satisfactoriamente estos datos masivos. El último periodo de "Auge" (2019-2022) se caracteriza por la consolidación y masificación de conceptos como aprendizaje de máquina, inteligencia artificial y aprendizaje profundo.

El uso de métodos estadísticos multivariados como el análisis de correspondencias y la descomposición de valores singulares en la recuperación de información de una colección de textos no estructurada, se basa en la propiedad contextual del lenguaje: palabras con significados semejantes, emergen en contextos similares. Dos o más palabras que tienen valores singulares semejantes, tendrán una connotación semántica similar y estarán siendo utilizadas en contextos también similares. Se infiere que dos o más documentos están refiriéndose a un tema similar, cuando sus valores singulares son parecidos. El método aquí presentado ha demostrado ser confiable para analizar grandes conjuntos de datos no estructurados, y por lo tanto relevar asociaciones latentes entre piezas lingüísticas.

Para calcular los pesos en la MDP se utilizaron las frecuencias relativas para cada documento (renglón) y para cada término (columna). Por lo tanto, este trabajo presenta una limitación en este sentido, dado que el método de asignación de pesos en función de las frecuencias relativas es el más sencillo de todos los existentes en la literatura. Se deja como trabajo futuro, la utilización de funciones asignación de pesos que han demostrado aumentar los niveles de precisión, de acuerdo con la literatura. Como por ejemplo la Función de Entropía (Phiri & Tiejun 2010) o la Función Frecuencia de Término – Frecuencia Inversa de Documento (TF-IDT), la cual es ampliamente discutida por Bafna, Pramod & Vaidya (2016).

El presente método está basado totalmente en los principios de la estadística multivariante, por lo tanto, resulta eficiente para automatizar la clasificación de documentos, o bien en otras tareas de procesamiento de texto que requieran de mínima supervisión. Esta característica también le hace susceptible de ser replicado para cualquier idioma o tema de investigación, sin necesidad de diccionarios o archivos auxiliares. Por otra parte, sus limitantes están relacionadas con la escalabilidad y rendimiento, dado que este método es computacionalmente muy costoso en comparación con otras técnicas de procesamiento de lenguaje natural.

## Referencias